\RequirePackage{ifpdf}
\documentclass[12pt]{JHEP3}         
\usepackage{amssymb,amsfonts}
\usepackage{cite}
\usepackage{amsmath}
\usepackage{graphicx}
\usepackage{MnSymbol}

\def\rmd{{\rm d}}

\newcommand{\nn}{\nonumber \\}

\newcommand{\ba}{\begin{eqnarray}}
\newcommand{\ea}{\end{eqnarray}}
\newcommand{\be}{\begin{equation}}
\newcommand{\ee}{\end{equation}}
\newcommand{\bal}{\begin{align}}
\newcommand{\eal}{\end{align}}
\newcommand{\bay}[1]{\left(\begin{array}{#1}}
\newcommand{\eay}{\end{array}\right)}

\newcommand{\Tr}{\mbox{Tr}}

\preprint{TAUP-3014/17}
\title{\center Supersymmetric R\'enyi entropy and Anomalies in $6d$ $(1,0)$ SCFTs}

\author{Shimon Yankielowicz\ ,~Yang Zhou\\

School of Physics and Astronomy, Tel-Aviv University\\ Ramat-Aviv 69978, Israel\\

{\tt E-mails : shimonya@post.tau.ac.il, yangzhou@post.tau.ac.il}
}

\abstract{A closed formula of the universal part of supersymmetric R\'enyi entropy $S_q$ for six-dimensional $(1,0)$ superconformal theories is proposed. Within our arguments, $S_q$ across a spherical entangling surface is a cubic polynomial of $\nu=1/q$, with $4$ coefficients expressed as linear combinations of the 't Hooft anomaly coefficients for the $R$-symmetry and gravitational anomalies. As an application, we establish linear relations between the $c$-type Weyl anomalies and the 't Hooft anomaly coefficients. We make a conjecture relating the supersymmetric R\'enyi entropy to an equivariant integral of the anomaly polynomial in even dimensions and check it against known data in $4d$ and $6d$.}

\begin{document}

\pagestyle{plain} \setcounter{page}{1}
\newcounter{bean}
\baselineskip16pt


\section{Introduction}
Six-dimensional superconformal theories provide a framework to understand various features of lower-dimensional supersymmetric dynamics.
By themselves, they are difficult to study by traditional quantum field theory techniques. All known examples of interacting CFTs in six dimensions are supersymmetric. The $(2,0)$ theories should be the simplest ones~\cite{Witten:1995zh,Strominger:1995ac,Witten:1995em}. A large class of interacting $(1,0)$ fixed points have been constructed in string theory or brane constructions~\cite{Seiberg:1996vs,Seiberg:1996qx,Ganor:1996mu,Blum:1997mm,Hanany:1997gh}. Recently, F-theory provides a way to classify the known and new $(1,0)$ fixed points~\cite{Heckman:2013pva,DelZotto:2014hpa,Heckman:2015bfa}.

Since all the known interacting fixed points are supersymmetric, it is expected that supersymmetry constraints are important in computing their physical characteristic quantities, such as Weyl anomalies. Indeed, the $a$-anomaly in $(1,0)$ superconformal theories has been recently determined in terms of their 't Hooft anomaly coefficients~\cite{Intriligator:2014eaa,Ohmori:2014kda} for the $R$-symmetry and gravitational anomalies~\cite{AlvarezGaume:1983ig} by analyzing supersymmetric RG flows on the tensor branch~\cite{Cordova:2015fha}~\footnote{The subscript $\mathfrak{u}(1)$ means an Abelian $(2,0)$ tensor multiplet. See~\cite{Cordova:2015vwa} for the result in $(2,0)$ theories and~\cite{Maxfield:2012aw} for earlier investigation.}
\be\label{normala}
\bar a = {a\over a_{\mathfrak{u}(1)}} = {16\over 7}\,(\alpha -\beta+\gamma) + {6\over 7}\,\delta\ ,
\ee
where $\alpha\ ,\beta\ ,\gamma\ ,\delta$ are the coefficients appearing in the anomaly polynomial
\be\label{anomalyP}
{\cal I}_8 = {1\over 4!} \left( \alpha\, c_2^2(R) + \beta\, c_2(R)p_1(T) + \gamma\, p_1^2(T) + \delta\, p_2(T)\right)\ .
\ee
Here $c_2(R)$ is the second Chern class of the $R$-symmetry bundle and $p_{1,2}$ are the Pontryagin classes of the tangent bundle. The relation (\ref{normala}) is analogues to the known relation~\cite{Anselmi:1997am} in four-dimensional ${\cal N}=1$ SCFTs,
$a_{d=4} = {9\over 32} k_{RRR} - {3\over 32} k_{R}$, where $k_{RRR}$ and $k_R$ are the $\Tr\,U(1)^3_R$ and $\Tr\,U(1)_R$ 't Hooft anomalies.
Although the anomaly multiplet in six dimensions has not yet been constructed, such linear relations are believed to follow from the anomaly supermultiplets which include 't Hooft anomalies as well as the anomalous trace of the stress tensor. The Weyl anomaly coefficients in $6d$ are defined from the latter~\cite{Deser:1976yx,Duff:1977ay,Fradkin:1983tg,Deser:1993yx}
\be\label{anomalousT}
\langle T_\mu^{\,\mu} \rangle \sim a\, E_6 + \sum_{i=1}^3 c_i\, I_i\ ,
\ee
where $E_6$ is the Euler density and $I_{i=1,2,3}$ are three Weyl invariants. In the presence of $(1,0)$ supersymmetry, $c_{i=1,2,3}$, satisfying a constraint $c_1-2c_2+6c_3=0$~\cite{Bastianelli:2000hi,deBoer:2009pn,Kulaxizi:2009pz}, are also believed to be linearly related the to 't Hooft anomaly coefficients~\cite{Beccaria:2015ypa,Butter:2016qkx,Butter:2017jqu}. Assuming that the linear relation indeed exist, one could determine its coefficients by considering the known values of the corresponding Weyl and 't Hooft anomalies in four independent examples. Unfortunately only three are known, i.e. the free hyper multiplet, the free tensor multiplet and supergravity~\cite{Bastianelli:2000hi,Bastianelli:1999ab}. The naive vector multiplet is not conformal and the conformal version~\cite{Beccaria:2015uta} involves higher derivatives. Evaluating the anomalies via the heat kernel method will involve higher powers of the Laplacian operator. We will have, therefore, to consider another approach. 

In even dimensions, it is known that the $a$-anomaly determines both the universal log divergence of the round-sphere partition function~\footnote{The $a$-anomaly is proportional to the coefficient of the log divergence.} and the universal log divergence in the vacuum state entanglement entropy associated with a ball in flat space~\cite{Casini:2011kv}. On the other hand, by the conformal Ward identities, the 2-point and 3-point functions of the stress tensor in the vacuum in flat space can be determined up to 3 coefficients~\cite{Osborn:1993cr,Erdmenger:1996yc}, which are linearly related to $c$-type Weyl anomalies $c_{1,2,3}$. In the presence of $(1,0)$ supersymmetry, only two of them are independent as mentioned before.

Because the round sphere is conformally flat, one expects that the nearly-round sphere partition function, which includes the response to a small deviation of the metric from the round sphere, is determined by the flat space stress tensor correlators. Due to these intrinsic relations and supersymmetric constraints, it is therefore tempting to ask whether one can fully determine the partition function on a branched ($q$-deformed) sphere,\footnote{A branched sphere is a sphere with a conical singularity with the deformation parameter $q-1$.} which is directly related to the supersymmetric R\'enyi entropy $S_q$.

Supersymmetric R\'enyi entropy was first introduced in three dimensions~\cite{Nishioka:2013haa,Huang:2014gca,Nishioka:2014mwa}, and later studied in four dimensions~\cite{Huang:2014pda,Zhou:2015cpa,Crossley:2014oea}, in five dimensions~\cite{Alday:2014fsa,Hama:2014iea}, in six dimensions ($(2,0)$ theories)~\cite{Nian:2015xky,Zhou:2015kaj} and also in two dimensions ($(2,2)$ SCFTs)~\cite{Giveon:2015cgs,Mori:2015bro}.
By turning on certain $R$-symmetry background fields, one can calculate the partition function $Z_q$ on a $q$-branched sphere $\Bbb{S}_q^d$, and define the supersymmetric R\'enyi entropy as \be\label{SREdefine}
S_q = {qI_1-I_q\over 1-q}\ , ~~~I_q:=-\log Z_q[\mu(q)]\ ,
\ee which is a supersymmetric refinement of the ordinary R\'enyi entropy (which is not supersymmetric because of the conical singularity).\footnote{For CFTs, the R\'enyi entropy (or the supersymmetric one) associated with a spherical entangling surface in flat space can be mapped to that on a sphere. Throughout this work we take the ``regularized cone'' boundary conditions.} The quantities defined in (\ref{SREdefine}) are UV divergent in general but one can extract universal parts free of ambiguities. 

\subsection{Summary of results}
The main result of this paper is the exact universal part of the supersymmetric R\'enyi entropy in $6d$ $(1,0)$ SCFTs. We show that, for theories characterized by the anomaly polynomial (\ref{anomalyP}), it is given by a cubic polynomial of $\nu={1/q}$
\be\label{closedF}
S^{(1,0)}_\nu = \sum_{n=0}^3 s_n (\nu-1)^n\ ,
\ee with four coefficients
\ba\label{closeds}
s_0 &=& {1\over 6}(8\alpha-8\beta+8\gamma+3\delta)\ ,\nn
s_1 &=& {1\over 4}(2\alpha-3\beta+4\gamma+\delta)\ ,\nn
s_2 &=& {1\over 24}(2\alpha - 5\beta + 8\gamma)\ ,\nn
s_3 &=& {1\over 192}(\alpha - 4\beta +16\gamma)\ .
\ea where $\alpha\ ,\beta\ ,\gamma\ ,\delta$ are the 't Hooft anomaly coefficients defined in (\ref{anomalyP}). The basic ingredients in our arguments are the following:

\begin{itemize}
\item[(A)] $S_\nu$ of $(1,0)$ free hyper multiplet and free tensor multiplet can be computed by the heat kernel method closely following~\cite{Zhou:2015kaj}. The results are given by
\ba
 S^{h}_\nu &=& \frac{7}{2880} (\nu -1)^3+\frac{7}{720} (\nu -1)^2+\frac{1}{40} (\nu -1)+\frac{11}{360}\ ,\label{freehyper0}\\
 S^{t}_\nu &=& \frac{1}{360} (\nu -1)^3+\frac{1}{90} (\nu -1)^2+\frac{1}{10} (\nu -1)+\frac{199}{360}\ .\label{freetensor0}
\ea These are the main results of Section \ref{SREfree}.

\item[(B)] $S_\nu$ of $A_{N-1}$ type $(2,0)$ theories (which are of course $(1,0)$ conformal theories) in the large $N$ has been computed in~\cite{Zhou:2015kaj}. The result is given by
\be
{S_\nu[A_{N\to\infty}]\over N^3} = \frac{1}{192} (\nu -1)^3+\frac{1}{12} (\nu -1)^2+\frac{1}{2} (\nu -1)+\frac{4}{3}\ .
\ee

\item[(C)] Based on (A)(B) and (F) below, a reasonable \underline{assumption} is that the general form of $S_\nu$ for $(1,0)$ SCFTs is a cubic polynomial in $\nu-1$. However, so far we do not have a sharp argument for this assumption.~\footnote{We are interested only in the universal part, i.e. the coefficient of the UV log divergent part. This part should be given by a finite number of counter-terms, each of them an integral of local functions of the supersymmetric background including the metric (squashed sphere). Unfortunately the supersymmetric smooth squashed sphere in $6d$ has not yet been constructed.}
Furthermore, based on (D)(E)(F) below, the four coefficients of the cubic polynomial are linear combinations of $\alpha, \beta, \gamma, \delta$.

\item[(D)] The value of $S_\nu$ at $\nu=1$ is the entanglement entropy associated with a spherical entangling surface, which is proportional to the $a$-anomaly (\ref{normala}).

\item[(E)] The first and second derivatives of $S_\nu$ at $\nu=1$ can be written as linear combinations of integrated two- and three-point functions of operators in supersymmetric stress tensor multiplet. Because of this, one can relate the first and second derivatives at $\nu=1$ to $c_1$ and $c_2$,
\be
\partial_\nu S_\nu \big|_{\nu=1} = {3\over 2}c_2-{3\over 4}c_1\ ,\quad \partial^2_\nu S_\nu \big|_{\nu=1}=c_2 - {5\over 16}c_1\ ,
\ee where $c_1$ and $c_2$ are believed to be given by linear combinations of 't Hooft anomaly coefficients $\alpha, \beta, \gamma, \delta$.

\item[(F)] The large $\nu$ behavior of $S_\nu$ is controlled by the ``supersymmetric Casimir energy''~\cite{Bobev:2015kza}. This gives
\be
\lim_{\nu\to\infty} {S_{\nu}\over \nu^3} = {1\over 192}(\alpha-4\beta+16\gamma)\ .
\ee

\item[(G)] In the large $\nu$ expansion, the second Pontryagin class (with coefficient $\delta$) will not contribute to the $\nu^3$ term (as we see from (F)) and the $\nu^2$ term. Because of the latter, one has
\be
\partial_\delta\left(\partial^2_\nu S_\nu \big|_{\nu=1}\right) = 0\ .
\ee

\item[(H)] For the conformal non-unitary $(1,0)$ vector multiplet, a constraint for the $c$-type Weyl anomalies, $c_1+4c_2={62\over 45}$, can be obtained by studying the higher-derivative operators on the Ricci flat background~\cite{Beccaria:2015ypa}.~\footnote{We thank Matteo Beccaria for explaining us this result first presented in~\cite{Beccaria:2015ypa}.} Together with (E), one has
\be
16 \left(\partial^2_\nu S^{\text{``Vector''}}_\nu \big|_{\nu=1}\right) - 8 \left(\partial_\nu S^{\text{``Vector''}}_\nu \big|_{\nu=1}\right)=(c_1+4c_2)\big|_{\text{``Vector''}}  = {62\over 45}\ .
\ee

\end{itemize} From (A)(B)(C)(D)(E)(F)(G)(H), one can uniquely find the general expression of the supersymmetric R\'enyi entropy given in (\ref{closedF})(\ref{closeds}). We emphasize that among all these ingredients (C) is an assumption, all the rest are derived results. The results (A),(D),(E),(F),(G) are new as far as we know. The precise agreement between (F) and (A)(B) can be considered as a nontrivial test of (F). Independently, we conjecture a relation between the supersymmetric R\'enyi entropy and the anomaly polynomial in any even dimension, which perfectly agrees with (A)-(H). We consider this precise agreement as a strong support of our result (\ref{closedF})(\ref{closeds}). Note that (E) and (\ref{closeds}) also establish the linear relations between $c$-type Weyl anomalies and the 't Hooft anomaly coefficients,\footnote{The numerical coefficients for the $c$-anomalies here are different from those presented in~\cite{Beccaria:2015ypa}, where an assumption concerning the structure of the linear combinations was made. We thank Matteo Beccaria for discussion on this issue. After our paper appeared on the arXiv the authors of~\cite{Beccaria:2015ypa} clarified to us that the data they used did not allow them to fix $c_{1,2,3}$ unambiguously. There was still a 1-parameter freedom consisting with our result. They fixed this freedom by another assumption/conjecture relating anomalies in $4d$ and $6d$.}
\ba\label{anomalyc}
c_1 &=& -{2\over 9}(6\alpha - 7\beta+8\gamma+4\delta)\ ,\nn
c_2 &=& -{1\over 18} (6\alpha - 5\beta + 4\gamma + 5\delta)\ ,\nn
c_3 &=& {1\over 18} (2\alpha - 3\beta + 4\gamma +\delta)\ .
\ea

This paper is organized as follows. In Section \ref{SREfree} we employ heat kernel method to study the supersymmetric R\'enyi entropy of free $(1,0)$ multiplets. In Section \ref{SREint} we propose a form of the universal supersymmetric R\'enyi entropy with four non-trivial coefficients, which works for general $6d$ $(1,0)$ SCFTs. We determine the coefficients one by one. We study the relation between the supersymmetric R\'enyi entropy and the supersymmetric Casimir energy in Section \ref{SREandE}, which is used to determine one of the coefficients in the previous section. In Section \ref{SREandP} we conjecture a relation between the supersymmetric R\'enyi entropy and the anomaly polynomial for SCFTs in even dimensions and test this conjecture in $6d$ and $4d$. In Section \ref{diss}, we discuss some open questions, further applications of our results and some future directions of research.

\section{Free $6d$ $(1,0)$ multiplets}
\label{SREfree}
We begin by studying the supersymmetric R\'enyi entropy of free $(1,0)$ multiplets, following~\cite{Nian:2015xky}. For free fields, the R\'enyi entropy associated with a spherical entangling surface in flat space can be computed by conformally mapping the conic space to a hyperbolic space $\Bbb{S}^1_\beta\times\Bbb{H}^5$ and using the heat kernel method.~\footnote{In this section we use $\beta=1/T$ as the inverse temperature and hopefully this will not be confusing with the anomaly coefficient $\beta$.} A six-dimensional $(1,0)$ hyper multiplet includes 4 real scalars, 1 Weyl fermion and a tensor multiplet includes 1 real scalar, 1 Weyl fermion and a 2-form field with self-dual strength. The 2-form field has a self-duality constraint which reduces the number of degrees of freedom by half.

\subsection{Heat kernel and R\'enyi entropy}
The partition function of free fields on $\Bbb{S}^1_{\beta=2\pi q}\times\Bbb{H}^5$ can be computed by the heat kernel~\footnote{For R\'enyi entropy of free fields in other dimensions less than six, see for instance~\cite{Casini:2010kt,Klebanov:2011uf,Fursaev:2012mp,Dowker:2012rp}.}
\be\label{partitionkernel}
\log Z(\beta) = {1\over 2} \int_0^\infty {dt\over t}K_{\mathbb{S}^1_\beta\times \mathbb{H}^5}(t)\ ,
\ee where $K_{\mathbb{S}^1_\beta\times \mathbb{H}^5}(t)$ is the heat kernel of the associated conformal Laplacian. The kernel factorizes because the spacetime is a direct product,
\be
K_{\mathbb{S}^1_\beta\times \mathbb{H}^5}(t)=K_{\mathbb{S}^1_\beta}(t)\, K_{\mathbb{H}^5}(t)\ .
\ee On a circle, the kernel is given by
\be
K_{\mathbb{S}^1_\beta}(t)={\beta\over \sqrt{4\pi t}}\sum_{n\neq 0,\in\mathbb{Z}} e^{-\beta^2n^2\over 4t}\ .
\ee
In the presence of a chemical potential $\mu$, it will be twisted~\cite{Belin:2013uta}
\be\label{twkernel}
\widetilde K_{\mathbb{S}^1_\beta}(t)={\beta\over \sqrt{4\pi t}}\sum_{n\neq 0,\in\mathbb{Z}} e^{{-\beta^2n^2\over 4t}+i2\pi n \mu+i\pi n f}\ ,
\ee where $f$ controls the periodic/anti-periodic boundary conditions, namely $f=0$ for bosons and $f=1$ for fermions. The volume factor can be factorized in the kernels on the hyperbolic space, because $\mathbb{H}^5$ is homogeneous. Thus $K_{\mathbb{H}^5}(t)$ can be written in terms of the equal-point kernel,
\be
K_{\mathbb{H}^5}(t) = \int d^5x \sqrt{g}~K_{\mathbb{H}^5}(x,x,t)=V_5\, K_{\mathbb{H}^5}(0,t)\ .
\ee The regularized volume is given by $V_5=\pi^2\log(\ell/\epsilon)$, where $\epsilon$ is actually the UV cutoff in the flat space before the conformal mapping~\footnote{In the replica trick approach to compute the entanglement/R\'enyi entropy, this is the $q$-fold space with a conical singularity.} and $\ell$ is the curvature radius of $\mathbb{H}^5$. For the $K_{\mathbb{H}^5}(0,t)$ of free fields with different spins we refer to~\cite{Nian:2015xky} and references there in.

The R\'enyi entropy of a hyper multiplet can be obtained by summing up the contributions of 4 real scalars, 1 Weyl fermion and the R\'enyi entropy of a tensor multiplet can be obtained by summing up the contributions of 1 real scalar, 1 Weyl fermion and a self-dual 2-form,
\ba
S^{hyper}_q &=& 4\times {S^s_q\over 2} + S^f_q\ ,\\
S^{tensor}_q &=& {S^s_q\over 2} + S^f_q + {S^v_q\over 2}\ .
\ea where the R\'enyi entropy for free fields with different spins can be computed by using the corresponding heat kernels.~\footnote{For some relevant details of this computation we refer to~\cite{Nian:2015xky}.} The final results for the R\'enyi entropy of a $6d$ complex scalar, a $6d$ Weyl fermion and a $6d$ $2$-form field are
\ba
S^s_q &=& \frac{(q+1) \left(3 q^2+1\right) \left(3 q^2+2\right)}{15120 q^5}{V_5\over \pi^2}\ ,\\
S^f_q &=& \frac{(q+1) \left(1221 q^4+276 q^2+31\right)}{120960 q^5}{V_5\over \pi^2}\ ,\\
S^v_q &=& \frac{(q+1) \left(37 q^2+2\right)+877 q^4+4349q^5}{5040 q^5}{V_5\over \pi^2}\ ,
\ea respectively.
Note that, to obtain the correct R\'enyi entropy for the two form field, we have taken a $q$-independent constant shift which is associated with possible boundary contributions~\cite{Nian:2015xky}.
Before moving on, let us represent $S_q$ in terms of 
$$S_{\nu}={\pi^2\over V_5}\, S_q\ ,~\text{with}~~ \nu=1/q\ .$$
The R\'enyi entropy of free $(1,0)$ multiplets are given by
\ba\label{tensorformula}
S^{hyper}_{\nu}&=&\frac{ (\nu -1)^5}{1920}+\frac{(\nu -1)^4}{320} +\frac{31(\nu -1)^3}{2880} +\frac{(\nu -1)^2}{45} +\frac{\nu -1}{30}+\frac{11}{360}\ ,\\
S^{tensor}_{\nu}&=&\frac{(\nu -1)^5}{1920} +\frac{(\nu -1)^4}{320} +\frac{13(\nu -1)^3}{960} +\frac{(\nu -1)^2}{30} +\frac{2(\nu -1)}{15}+\frac{199}{360}\ .
\ea The reason why $S_\nu$ is convenient is obvious, the series expansion near $\nu=1$ has finite terms while the expansion of $S_q$ near $q=1$ has infinite number of terms. We will use $S_{\nu}$ instead of $S_q$ to express R\'enyi entropy and supersymmetric R\'enyi entropy from now on.
It is worth to remember the relations between the derivatives with respect to $q$ and the derivatives with respect to $\nu$ at $q=1/\nu =1$,
\be\label{derivatives}
\partial_\nu S_\nu\big |_{\nu=1} = -\partial_q S_q\big |_{q=1}\cdot {\pi^2\over V_5} \ ,\quad \partial^2_\nu S_\nu\big |_{\nu=1} = \left(2\partial_q S_q+\partial^2_q S_q\right) \big |_{q=1}\cdot {\pi^2\over V_5}\ ,
\ee which will be useful later. Finally, one can check that $\partial^0_{q=1}$, $\partial^1_{q=1}$ and $
\partial^2_{q=1}$ of both $S^{hyper}_q$ and $S^{tensor}_q$ are consistent with the previous results  about the free $(1,0)$ multiplets~\cite{Bastianelli:2000hi,Bastianelli:1999ab}. By ``consistent'', we refer to the relations between the first and the second derivatives of the R\'enyi entropy at $q=1$ and the two- and three-point functions of the stress tensor derived in~\cite{Perlmutter:2013gua,Lee:2014zaa}.

\subsection{Supersymmetric R\'enyi entropy}
The supersymmetric R\'enyi entropy of free multiplets can be computed by the twisted kernel (\ref{twkernel}) on the supersymmetric background. The $R$-symmetry group of 6$d$ $(1,0)$ theories is $SU(2)_R$, which has a single $U(1)$ Cartan subgroup. Therefore one can turn on a single $R$-symmetry background gauge field (chemical potential) to twist the boundary conditions for scalars and fermions along the replica circle $\Bbb{S}_\beta^1$~\cite{Belin:2013uta}. The $R$-symmetry chemical potential can be solved by studying the Killing spinor equation on the conic space ($\Bbb{S}^6_q$ or $\Bbb{S}_{\beta=2\pi q}^1\times\Bbb{H}^5$),~\footnote{See the appendix in~\cite{Nian:2015xky}.}
\be\label{kschemical}
\mu(q) :=k\,A_\tau= {q-1\over 2}\ ,
\ee with $k$ being the $R$-charge of the Killing spinor under the Cartan $U(1)$. We choose $k=1/2$ and the background field turns out to be
\be\label{chemicalback}
A_\tau = (q-1)\ .
\ee
For each component field in the free multiplets, one has to first figure out the associated Cartan charge $k_i$ and then compute the chemical potential by $k_i\,A_\tau$. After that one can compute the free energy on $\Bbb{S}_\beta^1\times\Bbb{H}^5$ using the twisted heat kernel with the chemical potential $\mu = k_i\,A_\tau$ and obtain the supersymmetric R\'enyi entropy. 

After summing up the component fields, the supersymmetric R\'enyi entropy of a free $(1,0)$ hyper multiplet and a free $(1,0)$ tensor multiplet are
\ba
 S^{h}_\nu &=& \frac{7}{2880} (\nu -1)^3+\frac{7}{720} (\nu -1)^2+\frac{1}{40} (\nu -1)+\frac{11}{360}\ ,\label{freehyper}\\
 S^{t}_\nu &=& \frac{1}{360} (\nu -1)^3+\frac{1}{90} (\nu -1)^2+\frac{1}{10} (\nu -1)+\frac{199}{360}\ ,\label{freetensor}
\ea respectively.

\section{Interacting $6d$ $(1,0)$ SCFTs}
\label{SREint}
Having obtained the free multiplet results (\ref{freehyper})(\ref{freetensor}), we will use them to rewrite $S^{(1,0)}_\nu$ in a general form which, we hope, works for general interacting $6d$ $(1,0)$ SCFTs,
\be\label{ansatz}
S^{(1,0)}_\nu = A\, (\nu -1)^3+B\, (\nu -1)^2+C\, (\nu -1)+D\ ,
\ee
where the coefficients $A\,,B\,,C\,,D$ will depend on the specific theories.~\footnote{This structure is not true for the ordinary (non-supersymmetric) R\'enyi entropy~\cite{Galante:2013wta}.}

Before determining $A\,,B\,,C\,,D$ for general $(1,0)$ fixed points, let us summarize what we have learned so far for the existing examples.
These are free $(1,0)$ hyper multiplet, free $(1,0)$ tensor multiplet, $A_{N-1}$ type $(2,0)$ theories in the large $N$ limit and non-unitary conformal $(1,0)$ vector multiplet~\cite{Beccaria:2015ypa,Beccaria:2015uta}. We list $A\,,B\,,C\,,D$ and the relevant anomaly data for them in Table \ref{data}.~\footnote{We denote the conformal non-unitary vector multiplet by ``Vector''.} The anomaly data are from~\cite{AlvarezGaume:1983ig,Fradkin:1983tg,Bastianelli:2000hi}.

\begin{table}[htp]
\caption{Supersymmetric R\'enyi entropy and anomalies of known $(1,0)$ fixed points}
\begin{center}
\begin{tabular}{l*{11}{c}r}
$ $ & $A$ & $B$ & $C$ & $D$ & $\alpha$ & $\beta$ & $\gamma$ & $\delta$ & $c_1$ & $c_2$ & $c_3$\\
\hline
$\text{Hyper}$ & ${7\over 2880}$ & ${7\over 720}$ & ${1\over 40}$ & ${11\over 360}$ & $0$ & $0$ & ${7\over 240}$ & $-{1\over 60}$ & $-{1\over 27}$ & $-{1\over 540}$ & ${1\over 180}$ \\
$\text{Tensor}$ & ${1\over 360}$ & ${1\over 90}$ & ${1\over 10}$ & ${199\over 360}$ & $1$ & ${1\over 2}$ & ${23\over 240}$ & $-{29\over 60}$ & $-{8\over 27}$ & $-{11\over 135}$ & ${1\over 45}$\\
$A_{N-1} {1\over N^3}$ & ${1\over 192}$ & ${1\over12}$ & ${1\over 2}$ & ${4\over 3}$ & $1$ & $0$ & $0$ & $0$ & $-{4\over 3}$ & $-{1\over 3}$ & ${1\over 9}$\\
$\text{``Vector''}$ & $-$ & $-$ & $-$ & $-$ & $-1$ & $-{1\over 2}$ & $-{7\over 240}$ & ${1\over 60}$ & $-$ & $-$ & $-$ &\\
\end{tabular}
\end{center}
\label{data}
\end{table}

The coefficient $D$ in (\ref{ansatz}) can be determined by using the fact that, the entanglement entropy associated with a spherical entangling surface, which is nothing but $S_{\nu=1}$, is proportional to the Weyl anomaly $a$. This is true for general CFTs in even dimensions as shown in~\cite{Casini:2011kv}. Therefore
\be
{S^{(1,0)}_{\nu=1}\over S^{(2,0)}_{\nu=1}}={{a}\over a_{\mathfrak{u}(1)}} =: \bar a\ .
\ee
By studying supersymmetric RG flows on the tensor branch, $a/a_{\mathfrak{u}(1)}$ has been computed in~\cite{Cordova:2015fha}, see (\ref{normala}).
This allows us to fix
\be
D=S^{(1,0)}_{\nu=1}={7\over 12}\left({16\over 7}(\alpha-\beta+\gamma)+{6\over 7}\delta\right) = {4\over 3}(\alpha-\beta+\gamma) + {\delta\over 2}\ .
\ee

The coefficients $C$ and $B$ in (\ref{ansatz}) are the first and the second $\nu$-derivatives of $S^{(1,0)}_\nu$ at $\nu=1$, respectively. The transformations between the $\nu$-derivatives and the $q$-derivatives are given by (\ref{derivatives}). The relations between the $q$-derivatives and the integrated correlators are given in Appendix \ref{Pexp}.
Namely, the first $q$-derivative at $q=1$ is given by a linear combinations of integrated $\langle TT \rangle$ and integrated $\langle JJ \rangle$ in (\ref{susyprimeG}),
\be
S^\prime_{q=1} = -V_{d-1}\left({\pi^{{d\over 2}+1}\Gamma({d\over 2})(d-1)\over (d+1)!}C_T -g^2{\pi^{d+3\over 2}\over 2^{d-3}(d-1)\Gamma({d-1\over 2})}C_J\right)\ .
\ee
This relation holds for general SCFTs with conserved $R$-symmetry in $d$-dimensions.
Similarly the second $q$-derivative at $q=1$ is given by a linear combination of the integrated stress tensor 3-point function, the integrated $R$-current 3-point function and some mixed 3-point functions. This is given explicitly in (\ref{3pointc})
\be
S^{\prime\prime}_{q=1} ={1\over 6} I^{\prime\prime\prime}_{q=1}={4\pi^3\over 3}\left[\langle \hat E\hat E\hat E\rangle^c-g^3\langle \hat Q\hat Q\hat Q\rangle^c-3g\langle \hat E\hat E\hat Q\rangle^c+3g^2\langle \hat E\hat Q\hat Q\rangle^c\right]_{\Bbb{S}^1_{q=1}\times\Bbb{H}^{d-1}}\ .
\ee

In $6d$ $(1,0)$ SCFTs, by the conformal Ward identities, the two- and three-point functions of the stress tensor multiplet (including $R$-current) may be determined in terms of two independent coefficients, which are linearly related to $c_1$ and $c_2$. Because of this, $C$ and $B$ in (\ref{ansatz}) are also linear combinations of $c_1$ and $c_2$.
These relations can be obtained by fitting to the free hyper multiplet and the free tensor multiplet in Table \ref{data},
\be\label{BCc1c2}
B= c_2 - {5\over 16}c_1\ ,\quad C={3\over 2}c_2-{3\over 4}c_1\ .
\ee

Assuming $B$ and $C$ are linear combinations of $\alpha\,,\beta\,,\gamma\,,\delta$, we shall establish the explicit relations. Because the second Pontryagin class $p_2(T)$ does not contribute to the $\nu^2$ term, we get
\be
\partial_\delta B =0\ .
\ee To see that the $\nu^2$ term is independent of $p_2(T)$, let us consider the free energy on $\Bbb{S}^5_q\times\Bbb{H}^1$, which can be used to compute $S_q$ because $\Bbb{S}^5_q\times\Bbb{H}^1$ is conformally equivalent to $\Bbb{S}_q^6$ or $\Bbb{S}^1_q\times\Bbb{H}^5$. $\Bbb{S}^5_q\times\Bbb{H}^1$ is similar to $\Bbb{S}^5_q\times\Bbb{S}^1_{\beta\to\infty}$, but they are not the same due to different boundary conditions on $\Bbb{H}^1$ and $\Bbb{S}^1_\beta$. The latter background preserving supersymmetry is used to compute the supersymmetric Casimir energy in $6d$. One can formally define a supersymmetric R\'enyi entropy on $\Bbb{S}^5_q\times\Bbb{S}^1_{\beta\to\infty}$ with the R\'enyi parameter $q$ by using the free energy $\beta E_c[\Bbb{S}^5_q]$. As we will see in the next section, $p_2(T)$ will not contribute to the $1/q^2$ term in this supersymmetric R\'enyi entropy, because $p_2(T)$ contributes to $E_c$ in the following way (\ref{rule})
\be
{p_2(T)\over \omega_1\omega_2\omega_3} \to {1\over \omega_1\omega_2\omega_3} \sum_{i<j}^3\omega_i^2\omega_j^2\ ,\quad \omega_1=\omega_2=1\ ,\omega_3=1/q\ . 
\ee
The different boundary conditions on $\Bbb{S}^5_q\times\Bbb{H}^1$ will not change the property that the $1/q^2$ term is independent of $\delta$. We further confirm this fact by establishing a concrete relation between $S_q$ and the anomaly polynomial in Section \ref{SREandP}. 

Since $B$ depends only on $\alpha\,,\beta\,,\gamma\,$, it can be fixed by fitting to the three independent examples, the free hyper multiplet, the free tensor multiplet and the $A_{N-1}$ type theories in the large $N$,
\be\label{fixB}
B = {1\over 24} (2\alpha-5\beta+8\gamma)\ .
\ee
The same fitting method can be used to determine the $\alpha\,,\beta\,,\gamma\,,\delta$ dependence of $C$, but since $C$ depends on all four of them, one free parameter is left. We fix the remaining free parameter by making use of the result of $c_1+4c_2$ for the conformal non-unitary $(1,0)$ vector multiplet in~\cite{Beccaria:2015ypa} (obtained by the heat kernel computation on the Ricci flat background)
\be\label{c14c2}
(c_1 + 4 c_2)\big |_{\text{``Vector''}} ={62\over 45}\ .
\ee
Thus, the coefficient $C$ as a linear combination of $\alpha\,,\beta\,,\gamma\,,\delta$ is determined
\be\label{fixC}
C = {1\over 4} (2\alpha -3\beta + 4\gamma + \delta)\ .
\ee
(\ref{BCc1c2})(\ref{fixB})(\ref{fixC}) also establish the linear relations between $c_{1,2,3}$ and $\alpha\,,\beta\,,\gamma\,,\delta$
\ba
c_1 &=& -{2\over 9}(6\alpha - 7\beta+8\gamma+4\delta)\ ,\nn
c_2 &=& -{1\over 18} (6\alpha - 5\beta + 4\gamma + 5\delta)\ ,\nn
c_3 &=& -{1\over 6} (c_1-2c_2) = {1\over 18}(2\alpha -3\beta + 4\gamma +\delta)\ .
\ea
The remaining coefficient $A$ will be fixed as
\be
A={1\over 192}(\alpha - 4\beta + 16\gamma)\ .
\ee  in the next section by studying the large $\nu$ behavior of the supersymmetric R\'enyi entropy.
Obviously, the leading contribution in the limit $\nu\to\infty$ is determined only by $A$.

\subsection{A closed formula}
As a summary, we can completely determine a closed formula for the universal part of supersymmetric R\'enyi entropy for $6d$ $(1,0)$ SCFTs,
\ba\label{closedf}
S_\nu^{(1,0)} &=& {1\over 192}(\alpha - 4\beta +16\gamma) (\nu-1)^3+ {1\over 24}\,(2\alpha - 5\beta + 8\gamma)\,(\nu-1)^2\,\nn
&+& {1\over 4}(2\alpha-3\beta+4\gamma+\delta)(\nu-1)+{1\over 6}(8\alpha-8\beta+8\gamma+3\delta)\,.
\ea
Given that 't Hooft anomalies for general $6d$ $(1,0)$ SCFTs can be computed~\cite{Ohmori:2014kda}, the above formula tells us the universal supersymmetric R\'enyi entropy for any $(1,0)$ SCFT.

For $(2,0)$ theories labeled by a simply-laced Lie algebra $\mathfrak{g}$, (\ref{closedf}) reduces to~\cite{Zhou:2015kaj}
\be\label{alterF}
S^{(2,0)}_\nu = (\bar c-\bar a)\,{7\over 12}\, H_\nu + (7\bar a-4\bar c)\,{1\over 3}\, T_\nu\ ,
\ee where $\bar a$ and $\bar c$ are determined by the rank, dimension and dual Coxeter number of $\mathfrak{g}$,
\be 
\bar a = {16\over 7}\,d_\mathfrak{g}h^\vee_\mathfrak{g}  + r_\mathfrak{g}\ ,\quad \bar c = 4\,d_\mathfrak{g}h^\vee_\mathfrak{g} + r_\mathfrak{g}\ .
\ee
$T_\nu$ and $H_\nu$ are the supersymmetric R\'enyi entropy of the $(2,0)$ tensor multiplet and that of the $(2,0)$ supergravity (large $N$), respectively
\ba
T_\nu &=& \frac{1}{192} (\nu -1)^3+\frac{1}{48} (\nu -1)^2+\frac{1}{8} (\nu -1)+\frac{7}{12}\ ,\\
H_\nu &=& \frac{1}{192} (\nu -1)^3+\frac{1}{12} (\nu -1)^2+\frac{1}{2} (\nu -1)+\frac{4}{3}\ .
\ea

\section{Relation with supersymmetric Casimir energy}
\label{SREandE}
In this section we clarify the relation between the supersymmetric R\'enyi entropy and the supersymmetric Casimir energy in $6d$. Similar relation in $4d$ has been obtained in~\cite{Zhou:2015cpa}.
Recall that the partition function $Z$ on ${\cal M}^{D-1}\times\Bbb{S}^1_{\widetilde\beta}$ is determined by the Casimir energy on the compact space ${\cal M}^{D-1}$ in the limit $\widetilde\beta\to\infty$
\be
E_c := -\lim_{\widetilde\beta\to\infty}\partial_{\widetilde\beta} \log Z(\widetilde\beta)\ ,
\ee which is equivalent to the statement~\footnote{In this section we use $\widetilde\beta=1/T$ for the inverse temperature in order to distinguish it from the 't Hooft anomaly $\beta$.}
\be
\lim_{\widetilde\beta\to \infty}\log Z(\widetilde\beta) = -\widetilde\beta E_c\ .
\ee
We consider the cases with supersymmetry. In even-dimensional superconformal theories, the supersymmetric Casimir energy on $\Bbb{S}^1\times\Bbb{S}^{D-1}$ has been conjectured to be equal to the equivariant integral of the anomaly polynomial in~\cite{Bobev:2015kza}, where the authors provided strong supports for this conjecture by examining a number of SCFTs in two, four and six dimensions.~\footnote{For $6d$ superconformal index, see~\cite{Bhattacharya:2008zy,Kim:2012qf,Lockhart:2012vp}.} The equivariant integration is defined with respect to the Cartan subalgebra of the global symmetries (that commute with a given supercharge) and one can write this as
\be\label{equInteg}
E_D(\mu_j) = \int_{\mu_j} I _{D+2}\ ,
\ee 
where the equivariant parameters $\mu_j$ are the chemical potentials corresponding to the Cartan generators. In equivariant cohomology, doing the integration (\ref{equInteg}) in $6d$ is equivalent to the replacement rules (\ref{rule}).~\footnote{See the appendix in~\cite{Bobev:2015kza} for details on the equivalence.}

Let us consider $6d$ $(1,0)$ SCFTs on $\Bbb{S}^1_{\widetilde\beta}\times\Bbb{S}^5_{\vec\omega}$ with squashing parameters $\vec\omega = (\omega_1,\omega_2,\omega_3)$. The squashing parameters are defined by coefficients appearing in the Killing vector
\be
K = \omega_1{\partial\over\partial\phi_1} + \omega_2{\partial\over\partial\phi_2} + \omega_3{\partial\over\partial\phi_3}\ ,
\ee where $\phi_1,\phi_2, \phi_3$ are three circles representing the $U(1)^3$ isometries of the 5-sphere. The supersymmetric Casimir energy of superconformal $(1,0)$ theories is given by the equivariant integral (\ref{equInteg})
\be\label{casimirInteracting}
E^{(1,0)}_6(\mu_j) = -\int_{\mu_j} {\cal I}_8\ ,
\ee where the 8-form anomaly polynomial is~\footnote{We consider the minimal set of global symmetries without extra flavor symmetries.}
\be\label{anomalyP1}
{\cal I}_8 = {1\over 4!} \left( \alpha\, c_2^2(R) + \beta\, c_2(R)p_1(T) + \gamma\, p_1^2(T) + \delta\, p_2(T)\right)
\ee as introduced in the introduction. The integration (\ref{casimirInteracting}) is equivalent to the following replacement rules~\cite{Bobev:2015kza}
\be\label{rule}
c_2(R)\to -\sigma^2\ ,\quad p_1(T)\to \sum_{i=1}^3 \omega_i^2\ ,\quad p_2(T)\to \sum_{i<j}^3\omega_i^2\omega_j^2\ ,
\ee
 where $\sigma$ is the chemical potential for the $R$-symmetry Cartan and $\omega_{1,2,3}$ are the chemical potentials for the rotation generators (commuting with the supercharge). After the replacement, the result should be divided by the equivariant Euler class,
 \be
 e(T)=\omega_1\omega_2\omega_3\ .
 \ee
 In the particular background of $\Bbb{S}^5_q\times\Bbb{S}^1_{\widetilde\beta}$, where $\Bbb{S}^5_q$ is a $q$-deformed 5-sphere with the metric
 \be
 \rmd s^2 = (\sin^2\theta+q^2\cos^2\theta)\rmd\theta^2 + q^2\sin^2\theta\rmd\tau^2 + \cos^2\theta\rmd\Omega_3^2\ ,
 \ee
 one should identify the shape parameters as
\be
\omega_1=\omega_2=1\ ,\quad \omega_3={1\over q}\ .
\ee
Note that there is a supersymmetric constraint for the chemical potentials, $\sigma = {1\over 2}\sum_j\omega_j$.
Evaluating (\ref{casimirInteracting}) one obtains
\be
E^{(1,0)}_6 = -{1\over 24\omega_1\omega_2\omega_3}\left(\alpha\,\sigma^4 - \beta\,\sigma^2\sum_{j=1}^3 \omega_j^2 + \gamma\,\left(\sum_{j=1}^3 \omega_j^2\right)^2 + \delta\,\left(\sum_{i<j}^3\omega_i^2\omega_j^2\right)\right)\ .
\ee
Therefore the free energy in the $q\to 0$ limit~\footnote{$f:={I\over V}$, $I:=-\log Z$.}
\be
f[\Bbb{S}^5_{q\to 0}\times\Bbb{S}^1_{\widetilde\beta\to\infty}]={1\over \widetilde\beta\pi^2/2}\widetilde\beta E_c\bigg |_{q\to 0} =- {1\over 192\pi^2}{\alpha - 4\beta + 16\gamma\over q^3}\ ,
\ee where we have divided by a $q$-independent volume factor Vol $[\Bbb{D}^4\times\Bbb{S}^1_{\widetilde\beta}]=\widetilde\beta\pi^2/2$.
Because of the conformal equivalence between $\Bbb{S}^5_q\times\Bbb{H}^1$ and $\Bbb{S}^1_q\times\Bbb{H}^5$, we have
\be\label{fconnection}
f[\Bbb{S}^5_{q\to 0}\times\Bbb{S}^1_{\beta\to\infty}] = f[\Bbb{S}^5_{q\to 0}\times\Bbb{H}^1] = f[\Bbb{S}^1_{q\to 0}\times\Bbb{H}^5]\ ,
\ee where the first equality follows from the background coincidence and the second one follows from the conformal invariance of (supersymmetric) R\'enyi entropy and
\be
S_{q\to 0} = -I_{q\to 0}\ ,\quad I_q:=-\log Z_q\ .
\ee  From (\ref{fconnection}) we obtain the asymptotic supersymmetric R\'enyi entropy on $\Bbb{S}^1_q\times\Bbb{H}^5$
\be\label{renyiasympt}
S_{q\to 0} = -I_{q\to 0} = {1\over 192}{\alpha - 4\beta + 16\gamma\over q^3}\ .
\ee
This fixes the undetermined coefficient $A$ in (\ref{ansatz}) as
\be
A = {1\over 192}(\alpha - 4\beta + 16\gamma)\ .
\ee
Notice that this result perfectly agrees with the supersymmetric R\'enyi entropy of the known $(1,0)$ fixed points listed in Table \ref{data}. 

\section{Relation with anomaly polynomial}
\label{SREandP}
Inspired by the relation between the supersymmetric Casimir energy and the anomaly polynomial~\cite{Bobev:2015kza}, we conjecture in this section a relation between the supersymmetric R\'enyi entropy and the anomaly polynomial. Following this relation, the supersymmetric R\'enyi entropy in even dimensions can be extracted directly from the anomaly polynomial of the theory. We conjecture that $S_q$ is determined by an equivariant integral of the anomaly polynomial ${\cal I}_{D+2}$ with respect to the subalgebra formed by generators $(r, h_{j=1,...D/2}, h_{[{D\over 2}+1]})$, where $r$ is the $R$-symmetry Cartan generator and $h_j$ is the $j$-th orthogonal rotation generator in $\Bbb{R}^D$, while  $h_{[{D\over 2}+1]}$ generates an additional $U(1)$ rotation. We emphasize that we do not have yet a physical understanding of the extra $U(1)$, but just employ it in the same way as the other rotational $U(1)$'s. We will check our conjecture against existing data in $6d$ and $4d$. To simplify the notation, we will use $\widetilde h=h_{[{D\over 2}+1]}$ from now on. The Cartan generators commuting with a given supercharge $Q$ have the corresponding chemical potentials denoted by $\sigma,\vec\omega, \widetilde\omega$. Define an equivariant integral~\footnote{One can come up, for now, with some loose arguments that this equivariant integral gives the coefficient of the universal log divergence in the free energy on a general $D$-dimensional squashed sphere.}
\be\label{Ginteg}
F(\sigma,\vec\omega, \widetilde\omega) = \int_{(\sigma,\vec\omega, \widetilde\omega)} {\cal I}_{D+2}
\ee with the corresponding chemical potentials as the equivariant parameters. The supersymmetric R\'enyi entropy can be determined as follow
\be\label{SREcon}
S_q = V_{\Bbb{H}^1}{qF_1-F_q\over 1-q}\ ,\quad F_q=F(\sigma,\vec\omega, \widetilde\omega)\bigg|_{\vec\omega=\vec1,\widetilde\omega=1/q}\ .
\ee Note that in the second equation in (\ref{SREcon}), the supersymmetric constraint for the chemical potentials was implicitly assumed. A volume $V_{\Bbb{H}^1}=2\log(\ell/\epsilon)$ was factorized in $S_q$ because we work effectively on $\Bbb{S}_q^{D-1}\times\Bbb{H}^1$. We will test this conjecture for SCFTs in $4d$ and $6d$ in the following subsections. We have not been able, so far, to prove this conjecture. The fact that an equivariant integral appears in this conjecture may hint towards some localization.

\subsection{Six dimensions}
In $\Bbb{R}^6$, there is a $U(1)^3$ subalgebra in the rotation symmetries. The generators commuting with the supercharge have the corresponding chemical potentials, $\omega_{1,2,3}$. The additional chemical potential is $\widetilde\omega=\omega_4$. Consider superconformal theories with $SU(2)_R$ $R$-symmetry. For the 8-form anomaly polynomial given in (\ref{anomalyP1}), the replacement rule in carrying out the equivariant integration (\ref{Ginteg}) should be
\be\label{rule1}
c_2(R)\to -\sigma^2\ ,\quad p_1(T)\to \sum_{i=1}^4 \omega_i^2\ ,\quad p_2(T)\to \sum_{i<j}^4\omega_i^2\omega_j^2\ .
\ee After these replacements in the anomaly polynomial, we divide it by $\widetilde e(T)=\omega_1\omega_2\omega_3\omega_4$. The result is given by
\be
F(\sigma,\omega_{1,2,3,4}) = -{1\over 24\omega_1\omega_2\omega_3\omega_4}\left(\alpha\,\sigma^4 - \beta\,\sigma^2\sum_{j=1}^4 \omega_j^2 + \gamma\,\left(\sum_{j=1}^4 \omega_j^2\right)^2 + \delta\,\left(\sum_{i<j}^4\omega_i^2\omega_j^2\right)\right)\ .
\ee Upon plugging in
\be
\sigma = {1\over 2}\sum_{i=1}^4\omega_i\ ,\quad \omega_1=\omega_2=\omega_3=1\ ,\omega_4=1/q\ ,
\ee one obtains
\ba
{S_q\over \,V_{\Bbb{H}^1}} = {qF_1-F_q\over 1-q} &=&
\frac{\alpha -4 \beta +16 \gamma }{384 q^3}+\frac{13 \alpha -28 \beta +16 \gamma }{384 q^2}\nn&+&\frac{67 \alpha -76 \beta +112 \gamma +48 \delta }{384 q}+\frac{1}{384} (175 \alpha -148 \beta +112 \gamma +48 \delta).
\ea The above result can be rewritten as $S_\nu$,
\ba
S_\nu &=& \frac{1}{192} (\nu -1)^3 (\alpha -4 \beta +16 \gamma )+\frac{1}{24} (\nu -1)^2 (2 \alpha -5 \beta +8 \gamma )\nn
&+&\frac{1}{4} (\nu -1) (2 \alpha -3 \beta +4 \gamma +\delta )+\frac{1}{6} (8 \alpha -8 \beta +8 \gamma +3 \delta )\ .
\ea This agrees precisely with (\ref{closedf}). Remarkably, a single conjectured formula by the equivariant integral (\ref{SREcon}) can give the $a$-anomaly, $c_{1,2,3}$-anomalies and also a certain part of the supersymmetric Casimir energy simultaneously and precisely. We consider these agreements as a strong support of both our results (\ref{closedF}) and the conjecture itself.
\subsection{Four dimensions}
In $\Bbb{R}^4$, there is a $U(1)^2$ subalgebra in the rotation symmetries. The generators commuting with the supercharge have the corresponding chemical potentials, $\omega_{1,2}$. The additional chemical potential is $\widetilde\omega=\omega_3$. Consider superconformal theories with $U(1)_R$ $R$-symmetry. The 6-form anomaly polynomial is
\be
{\cal I}_6 = {1\over 3!}(k_{RRR}c_1(R)^3+k_{R}c_1(R)p_1(T))\ .
\ee
The supersymmetric Casimir energy is given by the equivariant integral of ${\cal I}_6$~\cite{Bobev:2015kza}
\be\label{Ec4d}
E_4 = \int {\cal I}_6 = {k_{RRR}\over 6\omega_1\omega_2}\sigma^3 - {k_R\over 24\omega_1\omega_2}(\omega_1^2+\omega_2^2)\sigma\ ,
\ee where the chemical potentials satisfy a supersymmetric constraint $\sigma={1\over 2}(\omega_1+\omega_2)$. Note that the relation between the conformal and the 't Hooft anomalies in a $4d$ ${\cal N}=1$ theory is
\be\label{kac}
k_{RRR}={16\over 9}\,(5a-3c)\ ,\quad k_R = 16\,(a-c)\ .
\ee Plugging this in (\ref{Ec4d}), one reproduces the familiar result~\cite{Assel:2015nca,Assel:2014paa}
\be
E_4 = {2\over 3}(a-c)(\omega_1+\omega_2) + {2\over 27}(3c-2a){(\omega_1+\omega_2)^3\over \omega_1\omega_2}\ .
\ee
For our purpose, the equivariant parameters have been generalized to $\sigma,\omega_1,\omega_2,\omega_3$. The equivariant integration (\ref{Ec4d}) now becomes
\be
F(\sigma,\omega_{1,2,3}) = {k_{RRR}\over 6\omega_1\omega_2\omega_3}\sigma^3 - {k_R\over 24\omega_1\omega_2\omega_3}(\omega_1^2+\omega_2^2+\omega_3^2)\sigma\ ,
\ee with a constraint $\sigma={1\over 2}(\omega_1+\omega_2+\omega_3)$. Evaluating the supersymmetric R\'enyi entropy (\ref{SREcon}), one obtains
\ba
S_\nu &=& {3\over 8} (k_R-3k_{RRR}) + \left({5\over 24}k_R - {3\over 8}k_{RRR}\right)(\nu-1) + {1\over 24}(k_R-k_{RRR})(\nu-1)^2\label{4dSREk}\ ,\\
&=&-4\,a-\frac{4}{3}\,c\,(\nu -1) - \frac{4}{27}\,(3c-2a)\,(\nu -1)^2\label{4dSRE}\ .
\ea
This is precisely the universal supersymmetric R\'enyi entropy in $4d$ ${\cal N}=1$. A few remarks are in order. The leading coefficient in large $\nu$, $-4(3c-2a)/27$, precisely agrees with the result in~\cite{Zhou:2015cpa}. The first $\nu$-derivative at $\nu=1$, $-4c/3$, agrees with (\ref{susyprimeG}).~\footnote{In the sense that $\partial_\nu S_{\nu=1}$ is a particular linear combination of $C_T$ and $C_J$, therefore proportional to $c$.} The constant term, $-4a$, agrees with the linear relation between the $a$-anomaly and the entanglement entropy in~\cite{Casini:2011kv}. From (\ref{4dSREk}) to (\ref{4dSRE}), we have used the relations (\ref{kac}). Demanding the equivalence between (\ref{4dSREk}) and (\ref{4dSRE}), one can reproduce the famous known relations between the conformal and the 't Hooft anomalies.

\section{Discussion}
\label{diss}
In this paper we proposed a closed formula for the universal log term of the six-dimensional supersymmetric R\'enyi entropy and made a conjecture that the supersymmetric R\'enyi entropy in even dimensions is equal to an equivariant integral of the anomaly polynomial. It remains a challenging problem to understand the extra $U(1)$ and to prove this conjecture. We leave it for future work.

Let us mention a few other open question and further directions of research that are related to this work.
\begin{itemize}
\item[1.] Proving our assumption that the expansion of the supersymmetric R\'enyi entropy in $1/q$ terminates (it is just a polynomial of $1/q$ with degree $3$ in $6d$). For this we need the dependence of possible counter-terms on $1/q$. Hence, we have to construct the six-dimensional supersymmetric curved background and in particular the smooth squashed six-sphere. The super-Weyl anomalies constructed on this background will give the universal part of the supersymmetric R\'enyi entropy. This approach will, hopefully, allow us to prove our assumption (C) in the introduction.

\item[2.] A generalization of the discussion in appendix A implies that the third derivative of the supersymmetric R\'enyi entropy is related to a specific linear combination of 4-point functions of the stress tensor and other operators in its multiplet. On the other hand, according to our result (\ref{closedF})(\ref{closeds}) it is related to $s_3$ and hence via (\ref{normala}) and (\ref{anomalyc}) to the Weyl anomalies. In $6d$, this is indeed consistent with a long time expectation that the $a$-anomaly should determine some specific term in the 4-point function of the stress tensor. This consistency becomes manifest for $(2,0)$ theories (\ref{alterF}). It would be nice to demonstrate the relation between $S_\nu^{\prime\prime\prime}|_{\nu=1}$ and the integrated 4-point functions of operators in the stress tensor multiplet in a straight forward way.

\item[3.] The supersymmetric R\'enyi entropy has been proven to satisfy the four inequalities~\cite{Zhou:2016kcz},
\be\label{Ineqs}
\partial_q S_q \leq 0\ ,\quad \partial_q\left({q-1\over q}S_q\right)\geq 0\ ,\quad \partial_q((q-1)S_q)\geq 0\ ,\quad \partial_q^2((q-1)S_q)\leq 0\ .
\ee
Imposing these information theory inequalities for the supersymmetric R\'enyi entropy, one can get bounds on the 't Hooft anomaly coefficients. For $4d$ ${\cal N}=1$ superconformal theories, plugging (\ref{4dSRE}) into (\ref{Ineqs}) one obtains ${3\over 7}\leq{a\over c}\leq{3\over 2}$. Notice that the lower bound is not as tight as the $4d$ ${\cal N}=1$ Hofman-Maldacena bounds. For $6d$ $(1,0)$ superconformal theories, plugging (\ref{closedf}) into (\ref{Ineqs}) one obtains~\footnote{It would be interesting to understand whether these bounds (or some of them) can be saturated by some specific theories.}
\ba
P_1:=\alpha-4(\beta-4\gamma)\geq 0\ ,\label{ITb1}\\
P_2:=3 \alpha -2 \beta \geq 0\ ,\label{ITb2}\\
67 \alpha -76 \beta +16 (7 \gamma +3 \delta )\geq 0\ ,\\
9 \alpha-8 (\beta -2 \gamma -\delta )\geq 0\ .
\ea
It is interesting to clarify the relations among different bounds in $6d$: the information theory bounds shown above, the unitary bound $C_T\propto c_3\geq 0$ which reads
\be
2 \alpha -3 \beta +4 \gamma +\delta \geq 0\ ,
\ee and the $6d$ supersymmetric Hofman-Maldacena bounds (obtained by free-multiplet estimaiton) in terms of $\alpha,\beta,\gamma,\delta$~\footnote{In terms of $c_1$ and $c_2$, the Hofman-Maldacena bound reads from Table \ref{data}, $20c_2\leq c_1\leq {40\over 11}c_2$. We thank Clay Cordova for telling us about the free-multiplet estimation approach.}
\ba
P_3:=8 \alpha -6 \beta +4 \gamma +7 \delta &\geq& 0 \label{HMb1}\\
P_4:=2 \alpha -9 \beta +16 \gamma -2 \delta &\geq& 0\ .\label{HMb2}
\ea It is interesting to notice that, $P_1$ is equal to $s_3$ in (\ref{closeds}) and, from $P_3\geq 0$ and $P_4\geq 0$ one can derive both $C_T\propto c_3\propto s_1\geq 0$ and $s_2\geq 0$.  This indicates that the inequalities $P_1\geq 0,P_2\geq 0,P_3\geq 0, P_4\geq 0$ are more fundamental than the others. Moreover, combining the information theory bounds (\ref{ITb1})(\ref{ITb2}) and the Hofman-Maldacena bounds (\ref{HMb1})(\ref{HMb2}), one obtains
\be
s_0={7\over 12}\bar a = {1\over 6}(8 \alpha -8 \beta +8 \gamma +3 \delta) = {1\over 48}(P_1+9P_2+4P_3+2P_4) \geq 0\ ,
\ee
which gives a proof of the positivity of the $a$-anomaly. We leave further investigation on different bounds for future work.
\end{itemize}
\section*{Acknowledgement}
We are grateful for helpful discussions with Ofer Aharony, Matteo Beccaria, Clay Cordova, Marcos Crichigno, Diego Hofman, Igor Klebanov, Zohar Komargodski, Ying-Hsuan Lin, Cobi Sonnenschein, Itamar Yaakov, Xi Yin and Kazuya Yonekura. YZ would like to thank the University of Amsterdam for hospitality. The work of S.Y. is supported in part by the Israel Science Foundation (ISF) Center of Excellence (grant 1989/14); by the US-Israel bi-national science foundation (BSF) grant 2012383 and by the German-Israel bi-national science foundation (GIF) grant I-244-303.7-2013. YZ is supported by ``The PBC program of the Israel council of higher education'' and in part by the Israel Science Foundation (grant 1989/14), the US-Israel bi-national fund (BSF) grant 2012383 and the German-Israel bi-national fund GIF grant number I-244-303.7-2013. 

\appendix

\section{Perturbative expansion around $q=1$}
\label{Pexp}
We review the perturbative expansion of supersymmetric R\'enyi entropy (associated with spherical entangling surface) around $q=1$. The great details have been given in~\cite{Zhou:2015kaj} and we will be brief.
Although our main concern will be $6d$ $(1,0)$ SCFTs, we keep the discussions valid for any SCFT with conserved $R$-symmetry in $d$-dimensions. 

Consider the supersymmetric partition function on $\Bbb{S}^1_{\beta=2\pi q}\times\Bbb{H}^{d-1}$ with $R$-symmetry background fields (chemical potentials),
\be
Z[\beta,\mu] = \Tr \left(e^{-\beta(\hat E-\mu \hat Q)}\right)\ .
\ee
which can be used to compute the supersymmetric R\'enyi entropy associated with a spherical entangling surface in flat space. We work with the grand canonical ensemble. The state variables can be computed as follows
\ba
E ~&=&~ \left({\partial I\over \partial\beta}\right)_\mu~ - {\mu\over\beta}\left({\partial I\over \partial\mu}\right)_\beta\ ,\label{Evariable}\\
S ~&=&~ \beta \left({\partial I\over \partial\beta}\right)_\mu~ - I\ ,\\
Q ~&=&~ -{1\over\beta}\left({\partial I\over \partial\mu}\right)_\beta\ ,\label{Qvariable}
\ea where $I:=-\log Z$.
The energy expectation value is given by (\ref{Evariable})
\be\label{Eexp}
E={\Tr (\rho\hat E) \over \Tr (\rho)}\ ,\quad \rho=e^{-\beta(\hat E-\mu \hat Q)}\ ,
\ee and the charge expectation value is given by (\ref{Qvariable})
\be\label{Qexp}
Q={\Tr (\rho\hat Q) \over \Tr (\rho)}\ .
\ee
In the presence of supersymmetry, both $\beta$ and $\mu$ are functions of a single variable $q$ therefore $I$ is considered as 
\be I_q:=I[\beta(q),\mu(q)]\ .
\ee The supersymmetric R\'enyi entropy is defined as
\be
S_q = {q I_1- I_q\over 1-q}\ .
\ee
Consider the Taylor expansion around $q=1$, with $\delta q:=q-1$,
\be\label{renyiexpansion}
S_q = S_{\text{EE}} + \sum_{n=2}^\infty{1\over n!}{\partial^n I_q\over\partial q^n}\bigg|_{q=1}\delta q^{n-1}\ .
\ee
The first $q$-derivative of $I_q$ is given by
\be
I^\prime_q = \left({\partial I\over \partial\beta}\right)_\mu\, \beta^\prime(q) + \left({\partial I\over \partial\mu}\right)_\beta\, \mu^\prime(q)\ .
\ee
Using (\ref{Evariable}) and (\ref{Qvariable}), one can rewrite it as
\be
I^\prime_q = (E-\mu Q)\, \beta^\prime(q) - \beta Q\, \mu^\prime(q)\ .
\ee
Plugging in the supersymmetric background,
\be\label{qbackground}
\beta(q) = 2\pi q\ ,\quad \mu(q)=g{q-1\over q}\ ,
\ee one finally has
\be\label{firstD}
I^\prime_q = 2\pi(E-g Q)\ .
\ee
Notice that $\mu(q)$ is solved from the Killing spinor equation. $g$ is some number depending on the $R$-charge of the preserved Killing spinor. In general both $E$ and $Q$ are functions of $q$. Moreover, $E$ and $Q$ here are expectation values rather than operators.

\subsection{$S^\prime_{q=1}$ and $I^{\prime\prime}_{q=1}$}
From (\ref{renyiexpansion}) we see that
\be
S^\prime_{q=1} = {1\over 2} I^{\prime\prime}_{q=1}\ .
\ee
Let us take one more $q$-derivative of (\ref{firstD}) and make use of (\ref{Eexp}) and (\ref{Qexp})
\be
I^{\prime\prime}_q = -4\pi^2\left( {\Tr \left(\rho(\hat E-g \hat Q)^2\right) \over \Tr (\rho)}-{\left[\Tr \left(\rho(\hat E-g \hat Q)\right) \right]^2\over \left[\Tr (\rho)\right]^2}\right)\ ,\label{sreprimemu}
\ee
which can be simplified by using $\rho_0=\rho(\mu=0)$ at $q=1$
\be\label{eqA16}
S^\prime_{q=1} = -2\pi^2 \left({\Tr \left(\rho_0(\hat E-g \hat Q)^2\right) \over \Tr (\rho_0)}-{\left[\Tr \left(\rho_0(\hat E-g \hat Q)\right) \right]^2\over \left[\Tr (\rho_0)\right]^2}\right)_{q=1}\ .
\ee
(\ref{eqA16}) can be written as connected correlators
\be
S^\prime_{q=1} = -2\pi^2 \left[\langle \hat E \hat E\rangle^c+g^2\langle \hat Q \hat Q\rangle^c-2g\langle \hat E \hat Q\rangle^c\right]_{\Bbb{S}^1_{q=1}\times\Bbb{H}^{d-1}}\ ,
\ee where we have used the fact that, $\hat Q$ is a conserved charge, $[\hat E,\hat Q]=0$, to flip the order of $\hat E$ and $\hat Q$. Given that $\langle \hat E \hat Q\rangle^c=0$ and $\langle \hat E \hat E\rangle^c$ has been computed in~\cite{Perlmutter:2013gua}, we get
\be\label{sreprime0}
S^\prime_{q=1} =-V_{d-1}{\pi^{d/2+1}\Gamma(d/2)(d-1)\over (d+1)!}C_T -2\pi^2 g^2\int _{\Bbb{H}^{d-1}}\int _{\Bbb{H}^{d-1}}\langle J_\tau(x)J_\tau(y)\rangle^c_{q=1}\ .
\ee $C_T$ is defined through the flat space correlator
\be
\langle T_{ab}(x)T_{cd}(0)\rangle = {C_T\over x^{2d}}I_{ab,cd}(x)\ ,
\ee where
\ba
I_{ab,cd}(x) &=& {1\over 2}\left(I_{ac}(x)I_{bd}(x)+ I_{ad}(x)I_{bc}(x)\right)-{1\over d}\delta_{ab}\delta_{cd}\ ,\nn I_{ab}(x)&=&\delta_{ab}-2{x_ax_b\over x^2}\ .
\ea
Now the task is to compute the second term in (\ref{sreprime0}). Following the way of computing $\langle TT\rangle$ on the hyperbolic space $\Bbb{S}^1_{q=1}\times\Bbb{H}^{d-1}$, one can make use of the flat space correlators in the CFT vacuum,
\be
\langle \hat Q \hat Q\rangle^c = -{\pi^{d-1\over 2}V_{d-1}\over 2^{d-2}(d-1)\Gamma({d-1\over 2})}C_J\ ,
\ee where $C_J$ is defined through the $R$-current correlator in flat space
\be
\langle J_a(x)J_b(0) \rangle = {C_J\over x^{2(d-1)}} I_{ab}(x)\ .
\ee
Our final result of $S^\prime_{q=1}$ becomes
\be\label{susyprimeG}
S^\prime_{q=1}=-V_{d-1}\left({\pi^{{d\over 2}+1}\Gamma({d\over 2})(d-1)\over (d+1)!}C_T -g^2{\pi^{d+3\over 2}\over 2^{d-3}(d-1)\Gamma({d-1\over 2})}C_J\right)\ ,
\ee
which shows that the first $q$-derivative of $S_q$ at $q=1$ is given by a linear combination of $C_T$ and $C_J$. This is intuitively expected because in the presence of supersymmetry, taking the derivative with respect to $q$ is equivalent to taking the derivative with respect to $g_{\tau\tau}$ and $A_\tau$ at the same time. 

In the particular case of $6d$ $(1,0)$ SCFTs, the 2-point function of the stress tensor is determined by the central charge $c_3$. Therefore the integrated 2-point function is proportional to $c_3$. Moreover, $S^\prime_{q=1}$ is also proportional to $c_3$, because the stress tensor and the $R$-current on the right hand side of (\ref{susyprimeG}) live in the same multiplet.
\subsection{$S^{\prime\prime}_{q=1}$ and $I^{\prime\prime\prime}_{q=1}$}
From (\ref{renyiexpansion}) we see that
\be
S^{\prime\prime}_{q=1} = {1\over 6} I^{\prime\prime\prime}_{q=1}\ .
\ee
It is straightforward to compute $I^{\prime\prime\prime}_{q}$ by taking one more derivative on (\ref{sreprimemu})
\ba
{I^{\prime\prime\prime}_q\over 8\pi^3} &=& {\Tr \left(\rho(\hat E-g \hat Q)^3\right) \over \Tr (\rho)}-3{\Tr \left(\rho(\hat E-g \hat Q)^2\right)\Tr\left(\rho(\hat E-g \hat Q)\right)\over \left[\Tr (\rho)\right]^2}\nn
&&~~+2{\left[\Tr \left(\rho(\hat E-g \hat Q)\right) \right]^3\over \left[\Tr (\rho)\right]^3}\ ,
\ea which may be simplified at $q=1$ where $\mu=0$
\ba
{I^{\prime\prime\prime}_{q=1}\over 8\pi^3} &=& \bigg({\Tr \left(\rho_0(\hat E-g \hat Q)^3\right) \over \Tr (\rho_0)}-3{\Tr \left(\rho_0(\hat E-g \hat Q)^2\right)\Tr\left(\rho_0(\hat E-g \hat Q)\right)\over \left[\Tr (\rho_0)\right]^2} \nn &+& 2{\left[\Tr (\rho_0(\hat E-g \hat Q))\right]^3\over \left[\Tr (\rho_0)\right]^3}\bigg)_{q=1}\ .
\ea This can be further written in terms of connected correlation functions,
\be\label{3pointc}
S^{\prime\prime}_{q=1} ={1\over 6} I^{\prime\prime\prime}_{q=1}={4\pi^3\over 3}\left[\langle \hat E\hat E\hat E\rangle^c-g^3\langle \hat Q\hat Q\hat Q\rangle^c-3g\langle \hat E\hat E\hat Q\rangle^c+3g^2\langle \hat E\hat Q\hat Q\rangle^c\right]_{\Bbb{S}^1_{q=1}\times\Bbb{H}^{d-1}}\ ,
\ee where we have used $[\hat E,\hat Q]=0$. The integrated correlators in (\ref{3pointc}) can be computed by transforming the corresponding flat space correlators, $\langle TTT\rangle, \langle JJJ\rangle, \langle TTJ\rangle, \langle TJJ\rangle$ in the CFT vacuum. These correlators in flat space can be determined up to some coefficients for $d$-dimensional CFTs by conformal Wald identities~\cite{Osborn:1993cr,Erdmenger:1996yc}. In the case of $6d$ $(1,0)$ SCFTs, both the 2- and 3-point functions of the stress tensor multiplet can be determined in terms of three coefficients $c_{1,2,3}$.~\footnote{$c_3$ is not independent.} Therefore the right hand side of (\ref{3pointc}) should be proportional to some linear combinations of $c_{1,2,3}$, because the stress tensor and the $R$-current belong to the same multiplet.

\end{document}